# Thermochemistry of $MgB_2$ Thin Film Synthesis

Jihoon Kim, Rakesh K. Singh, Nathan Newman, and John M. Rowell

**Abstract— We have investigated the thermodynamic and kinetic barriers involved in the synthesis of $MgB_2$ films. This work refines our initial conjectures predicting optimal $MgB_2$ thin film growth conditions as a consequence of the unusually large kinetic barrier to $MgB_2$ decomposition. The small Mg sticking coefficient at temperatures greater than 300°C prevents high temperature synthesis with traditional vacuum growth methods. However, as a result of the large kinetic barrier to $MgB_2$ decomposition, *in-situ* thermal processing can be used to enhance the crystallinity and the superconductivity of $MgB_2$ films. We used these methods to produce $MgB_2$ thin films with relatively high transition temperatures (~37K) by pulsed laser deposition (PLD).**

## I. INTRODUCTION

A transition temperature Tc of 39K[1] makes $MgB_2$ an attractive candidate material for use in digital superconducting circuits operating at temperatures up to 25K, as well as in a number of other superconducting applications. The first step in the development of a technology for all electronics applications is the synthesis of thin films of $MgB_2$ with properties as close as possible to those of the best bulk samples. Ideally, it would be preferable to develop an as-made in-situ or one-step in-situ deposition process, i.e., one in which there is no anneal following the deposition. It has become clear that $MgB_2$ film synthesis strikingly illustrates the importance of understanding and controlling the condensation and desorption of the Mg and B species (which define their sticking coefficients), their surface diffusion and nucleation, bulk diffusion, and the $MgB_2$ phase formation (and the formation of competing phases, including impurity phases such as MgO), as well as decomposition at both the growth temperature and at post-anneal temperatures, if such anneals are used. In this paper, we will describe our present understanding of these issues, and we will illustrate many of them by presenting the results of our own film growth studies.

## II. EXPERIMENTAL METHOD

Our deposition system is built to ultra high vacuum specifications but so far has not been baked to improve its background pressure. It is pumped by a cryopump (CTI model 8). Before deposition, the typical pressure is $10^{-8}$ torr, during deposition it rises to ~$10^{-7}$ torr. Deposition of the B is by laser ablation from a B target of 99.9% purity. A 248 nm KrF excimer laser is used at a repetition rate of 8 Hz. Mg is evaporated from a Knudsen cell through a 2.5 cm diameter 25 cm long cylindrical cryoshroud cooled by liquid nitrogen. The flux of B and Mg is measured with a water-cooled quartz crystal monitor. 1 $cm^2$ sapphire substrates are fixed with silver paste to a resistively heated substrate that can reach 900°C. Electrical properties were characterized using a dipping probe with spring loaded contacts. The thickness of the films, and their composition, was determined by the routine use of Rutherford Backscattering.

## III. RESULTS AND DISCUSSION

### A. Evaporation of Mg from Mg

The vapor pressure of Mg over Mg metal is taken from tables and is reproduced in Fig. 1. We have also included another scale that shows the evaporation rate, in Angstroms per second. As is well known, Mg has extremely high vapor pressures at relatively low temperatures, for example $10^{-7}$ torr at 210°C, increasing to 1 torr at 600°C. As the melting point of Mg is 649°C, generally Mg vapor sublimes from the solid, rather than evaporating from the melt. To illustrate the implication of such high Mg vapor pressures for $MgB_2$ growth, let us assume that we wish to grow an $MgB_2$ film of thickness 2000 Å in a deposition time of 20 minutes, or at a deposition rate of 6000 Å/hour, or 1.7 Å/s. We can see in Fig. 1 that, if the substrate temperature is 270°C, Mg will be lost at this rate. Hence, unless Mg reacts rapidly with B to form $MgB_2$ at this temperature, there will be no net accumulation of Mg on the substrate at this or at higher temperatures, unless the Mg incorporation rate (i.e. product of impinging flux and sticking coefficient) is above 1.7 Å/s.

### B. Evaporation of Mg from $MgB_2$

Given the known high vapor pressures of Mg over Mg metal, it seemed important to us early in 2001 to determine the vapor pressure of Mg over $MgB_2$, and hence the evaporation rate of Mg from $MgB_2$. The higher this evaporation rate, the more difficult it is to attain useful growth rates at substrate

Manuscript received August 6, 2002. This work was supported by the Office of Naval Research (Contract No. N00014-02-1-0002).

J. Kim is with the Chemical and Materials Engineering Department, Arizona State University, Tempe, AZ 85287 USA (phone: 480-727-6934; fax: 480-965-0037; e-mail: Jihoon.Kim@asu.edu).

R. K. Singh is with the Chemical and Materials Engineering Department, Arizona State University, Tempe, AZ 85287 USA. (e-mail: Rakesh.K.Singh@asu.edu).

N. Newman is with the Chemical and Materials Engineering Department, Arizona State University, Tempe, AZ 85287 USA. (e-mail:Nathan.Newman@asu.edu).

J. M. Rowell is with Northwestern University, Evanston, IL 60208 and visiting Professor, Arizona State University, Tempe, AZ 85287 USA. (e-mail:jmrberkhts@aol.com).

temperatures which form a well ordered crystalline phase. Our measurements of the Mg vapor pressure over a bulk source of $MgB_2$ were reported earlier[2] and the results are presented in Fig. 1. For comparison, we also show the expected vapor pressure calculated by Z.-K. Liu et al from the thermodynamics of the $MgB_2$ system[3]. It is immediately clear that the vapor pressure we measured is significantly lower than that expected thermodynamically, by a factor of over $10^3$. This small evaporation coefficient of $MgB_2$ implies, for example, that if a growth temperature of 650°C is selected, the loss of Mg from the compound film will be at a rate of 1 monolayer/second, whereas $10^3$ monolayers/second, or about 3,000Å/second, would have been expected from thermodynamic considerations alone.

### C. Sticking probability of Mg

We can study the sticking probability of Mg and B by reference to the results of our as-made film growth studies, which are summarized in Fig. 2(a). Depositions were made at a flux of Mg, set by the quartz crystal monitors at 1 Å/sec. At $T_{dep}$=150°C, we obtained a film thickness of 2100Å for a deposition of 40 minutes, or a deposition rate of ~ 3200 Å/hour. At 200°C, this deposition rate has dropped to 800Å/hour, and at 250 °C to 400°C it is in the range of 200 to 300Å/hour, for the same incident Mg flux. The film stoichiometry also changes with $T_{dep}$, as shown in Fig. 2(b). At 150°C, the Mg:B ratio is about 1.3:2, at 200 and 250°C it is very close to 1:2, at 300 and 350°C it has dropped to about 1:4, while at 400°C it is about 1:5. The rapid decrease in thickness shows that, for temperatures above 250°C, the film growth rate has declined to less than one tenth of its value at 150°C. We have already shown that this is not due to the decomposition of fully reacted $MgB_2$. Rather, it is due to a very low sticking probability of Mg onto the substrate and onto the deposited B. The sticking coefficients of Mg depending on temperature and flux are shown on Fig. 2(a). Above 200°C, the Mg sticking coefficient becomes lowered to less than 0.17, from 0.74 at 150°C, when the Mg flux is 1 Å/sec. Though higher flux rate surely enhances the sticking coefficient at the same temperature, the increase in the sticking coefficient at higher flux becomes less at higher temperatures. On the other hand, boron shows no change in its sticking coefficient as temperature changes. These results illustrate the key problem in $MgB_2$ film growth in high vacuum, namely that low sticking probabilities imply that there is only a small probability that Mg reacts with B to form a non-volatile Mg-B complex (e.g. $MgB_2$) on the surface. Therefore, very little film growth is accomplished above 200°C, unless the Mg flux is increased to very high levels to enhance the reaction probability with the surface B species. Fig. 2 shows that enhanced Mg rates result in increases in the Mg:B ratio to over 1:2, as well as the Tcs. The enhanced rate of Mg, which is almost certainly the faster diffusing species on the growth surface, would be expected to enhance the effective surface diffusion rate, analogous to group-III rich growth in III-N MBE growth, as well as to help drive the synthesis reaction. Therefore enhanced Mg rates would presumably lead to improved $MgB_2$ crystallinity and higher Tcs, as is experimentally found. Our findings are also consistent with the formation of a high-resistive non-superconducting meta-stable phase for sub-stoichiometric $MgB_2$ compounds [4].

For temperatures greater than ~220°C, when the evaporation rate from Mg is greater than the Mg incorporation rate (left of the intersection of these curves in fig. 1), the Mg remaining in the film must exist in non-volatile Mg-B compounds. Note that the sticking coefficient exhibits a measurable and gradual drop at temperatures higher than where all the Mg metal would have been desorbed (i.e. > ~240°C), but lower than where significant $MgB_2$ decomposition begins. This indicates that Mg must form an intermediate Mg-B chemisorbed species with a bond strength that falls between that of Mg metal and fully reacted $MgB_2$ compounds.

### D. $MgB_2$ phase formation

An approximate rule of thumb, often used to estimate the substrate temperatures required to develop high-quality epitaxial films, is that the deposition temperature should be between one half and two thirds of the melting point. As $MgB_2$ melts at 2700K, this would mean that films should be grown at temperatures above 1000°C. Fortunately, many films grow highly-oriented, and even epitaxial, at much lower temperatures, and this is the case for $MgB_2$. Ueda and Naito[5] showed that films with Tc$_{onset}$ and Tc$_{zero}$ of 12.2 K and 5.2K, respectively, could be made by growth at 150°C. [Henceforth, we use 12.2/5.2K, to denote Tc$_{onset}$/Tc$_{zero}$.] Our films at 150°C were not superconducting, but at 200°C the Tc was 13.9/12.4K for our as-made films produced with the higher Mg impingement rate (1.5 Å /s). It is interesting to note that films deposited at 200 and 250°C with a Mg:B impinging flux ratio of 1:2 (i.e. a Mg flux of 1 Å /s) did not exhibit superconductivity, presumably as a result of poor crystallization and small grain size.

Our results, and those of Ueda and Naito, indicate that substrate temperatures of only 150 to 200°C are adequate to produce some $MgB_2$ phase in the film. Whether such low deposition temperatures can result in well-ordered films, with resistivities comparable to low bulk values (10μΩ-cm at 300K)[6], remains to be seen. Determination of the resistivity of as-made films is sometimes affected by the presence of excess unreacted Mg. In our deposition system, we can view the film at it is growing. Films made at ~200-225°C appear matt and gray, but on annealing above 500°C, quite quickly become highly reflective. We believe this is a direct observation of the evaporation of excess Mg. Thus it is likely that resistivity measurements of films made at relatively low temperatures can be affected by excess Mg.

### E. Temperature window for in-situ film growth

We can summarize our discussion of the constraints on $MgB_2$ film growth in vacuum. The small sticking probability of Mg at temperatures above 325°C limits the deposition temperature to the range below this value, unless the Mg flux is increased to extremely high values. Below 300°C, the crystalline order of $MgB_2$ appears to be limited. The decomposition of $MgB_2$, and the evaporation of Mg from $MgB_2$, limits anneal temperatures in vacuum to temperatures below about 850°C. Consideration of these constraints defines the various modes of $MgB_2$ film growth: (a) As-made growth at temperatures below 325ºC, (b) As-made growth at temperatures above 325 ºC with high Mg fluxes in vacuum, (c) As-made growth at temperatures above 325ºC by processes employing high Mg-containing gas pressures, such as MOCVD, (d) Deposition at temperatures below 325ºC, followed by an in-situ anneal at high temperatures typically in a high pressure of inert gas.

*F. Growth of $MgB_2$ films – as-made, heat-cycle, and in-situ post anneal.*

We have grown $MgB_2$ films by three processes - as-made, by what we term a heat cycle process, and using an in-situ post-anneal in argon.

(a) As-made (single step growth): Using an as-made process, we were successful in obtaining superconducting films at deposition temperatures of 200ºC (13.9/12.4 K) and at 225ºC (14.3/11.7K). To obtain superconducting films, our incident flux of Mg was increased to 1.5 Å/sec at 200ºC, and to 3 Å/sec at 225ºC. Our RBS measurements show that the Mg:B composition is close to 1:2 up to 250ºC. Fig. 1 shows the decrease in the growth rate due to the diminishing Mg sticking coefficient. However, there was no significant decomposition during the single step process. The growth rate is orders of magnitude faster than the $MgB_2$ decomposition rate in this low temperature range as a result of the relatively low growth temperature.

(b) Heat-cycle process: This process was an attempt to simulate a series of in-situ thermal anneals on effectively thin layers. The films were grown following a cycle of 10 minutes of deposition at a substrate deposition temperature ($T_{dep}$) of ~200ºC-225ºC, followed by increasing the substrate temperature to a higher ultimate temperature, designated $T_{max}$, at a ramp rate of about 120ºC/minutes. As soon as $T_{max}$ was reached, the heater current was decreased, and the temperature returned to $T_{dep}$ at about 40ºC/minutes. This cycle was repeated 5 times, with the Mg and B fluxes remaining fixed. We hoped that the $MgB_2$ phase would form at the relatively low deposition temperature, and that crystalline order would be enhanced during the cycle to higher anneal temperatures, without loss of Mg from the $MgB_2$, as the values of $T_{max}$ were kept below 600ºC. Tcs of these heat-cycle films are shown in Fig. 3 and Fig. 5(a). This figure reveals the following trends. The film thickness, which is 1750Å for the initial deposition at 200ºC (with excess Mg), decreases steadily as $T_{max}$ increases, dropping to about 500Å for $T_{max}$= 550ºC. During the cycled growth, 50 minutes is spent at 200ºC, and, as the films are superconducting as-made at 200ºC, clearly some $MgB_2$ phase must be formed. For temperatures above ~ 250ºC, all of the free Mg should be desorbed. Significant amounts of additional Mg should not be lost from $MgB_2$ until we reach $T_{max}$ of more than 500ºC. Perhaps much of the film at 250ºC and higher must still be in the form of an intermediate Mg-B chemisorbed species, which is lost during the thermal cycles. Clearly, the bond strength of this species must fall between the relatively strongly bound $MgB_2$ and the weakly bound free Mg metal. Overall, the Mg:B stoichiometry falls at or below 1:2. It is also possible that Mg-rich regions with species such as free

Mg or $Mg_3B_2$ are buried below the surface and therefore only susceptible to desorption after exposure to sufficient time and temperature. However, we find, that by comparing the chemical composition of films exposed to a number of heats cycles, with a single anneal (as described in the next section), that the influence of this effect is not the rate-limiting step.

The Mg:B stoichiometry, which is over 1:1 at 200ºC, also drops, becoming 1:2 at $T_{max}$=300 and 350ºC, but only 1:4 at 550ºC. Figure 5 illustrates the electrical properties of the films as a function of temperature. The Tc, which is 13.9/12.4K without any anneal, drops to 10.0/6.0K for $T_{max}$ of 300ºC, then slowly increases to 21.0/14.4K at $T_{max}$ of 550ºC. The high resistivities (factor of 10 and higher) yet similar form of the temperature dependence of the resistivity, when compared to low resistivity bulk samples[6], presumably indicates that these films form a loosely connected network of superconducting regions with reduced effective cross-sectional area. The reduced Tc indicates that these films are in the dirty limit as a result of poor crystallinity and/or small grain sizes.

(c) In-situ post-anneal: Our in-situ anneal process begins with a similar deposition at 225ºC. As soon as the deposition is complete, the chamber is filled with 10 torr argon, and the heater temperature is increased to an ultimate temperature, $T_{max}$. In some cases, the heater power was turned off as soon as $T_{max}$ was reached, in other cases the anneal temperature was maintained for times as long as 30 minutes. The properties of our in-situ post-annealed films are summarized in Fig. 4 and Fig. 5. As we noted earlier, the films made at 225ºC have excess Mg and are nearly 2000Å thick. As the anneal temperature is increased, the thickness drops as free Mg or weakly bound Mg evaporates away, and is in the range between 250 and 550Å for all the annealed films. The Mg:B stoichiometry is 1:2 for $T_{max}$ from 650 to 750ºC, then drops to 1:2.5 by 850ºC. The transition temperatures of the films are increased substantially by the anneals. We note that using our heat-cycle process to 550ºC, we only achieved a Tc of 21.0/14.4K. However, the post anneal in argon yields films with Tcs of 27.5/25.5K if a $T_{max}$ of 550ºC is maintained for only 10 seconds, whereas 33.5/31.5K results from an anneal at 550ºC for 30 minutes. For short anneals, the maximum Tcs are achieved at 750 and 800ºC, being 35/33.5K. Fig. 6 shows the decomposition rates during heat-cycle and post-anneal processes. The figure clearly illustrates similar net Mg loss occurs for both processes, at least at the common thermal processing temperature of 550ºC. Our studies clearly show that a single post-anneal results in higher quality films and higher Tcs than the multiple step heat cycling. It appears that the single annealing step process on the thicker layers and the heat cycling process on effectively thinner layers result in similar chemical stoichiometries for similar annealing temperatures and total times. The role of the thicker layers appears to produce improved electrical properties via improved crystalinity and grain size. TEM studies are in progress to experimentally determine the resulting differences in microstructure.

This post anneal scheme was also applied to various substrates such as MgO, AlN(2000Å)/SiC, etc. These films were grown with the same condition, deposition at 225ºC and anneal at 700ºC for 1 minute. Fig. 5 includes a 37K $Tc_{onset}$ of the film grown on an AlN buffer layer. We will publish details of the role that the substrate has on the chemical and electrical properties of $MgB_2$ thin films in a future publication.

## IV.  CONCLUSION

The small Mg sticking coefficient limits the initial growth temperature to less than 300°C for $MgB_2$ vacuum deposition methods. The large kinetic barriers to the decomposition of $MgB_2$ makes it possible to use in-situ thermal processing at elevated temperatures to synthesize high quality $MgB_2$. With this approach, we have produced $MgB_2$ with 37K $Tc_{onsets}$ using pulsed laser deposition.

## ACKNOWLEDGEMENTS

We would like to thank Deborah Van Vechten for her encouragement and support, as well as Barry Wilkens for assistance in the RBS measurements and John Kouvetakis and Lei Yu for helpful discussions.

# Figure Captions

Fig. 1.  Comparison of the Mg vapor pressure over Mg metal and $MgB_2$.  Comparison is made with thermodynamic predictions of Reference 3.   The Mg incorporation rate during PLD growth is found to diminish as elevated temperatures as a result of a reduced sticking coefficient. Note that above $220^oC$, the Mg evaporation rate from metallic Mg is greater than the Mg incorporation rate achieved during our PLD growth.

Fig. 2.  Growth rate, sticking coefficient, and stoichiometry of $MgB_2$ thin films synthesized with various Mg fluxes by PLD. The reduced sticking Mg coefficient results in the reduced growth rates and $Mg:B_2$ ratios with increasing temperature.

Fig. 3.  Summary of the heat-cycle process, (a) Tc and (b) thickness and stoichiometry of the deposited $MgB_2$ films.

Fig. 4.  Summary of the in-situ post-anneal process, (a) Tc and (b) thickness and stoichiometry of the deposited $MgB_2$ films.

Fig. 5.  Electrical properties of $MgB_2$ (a) formed by the heat-cycle process on sapphire and (b) formed by the post-anneal process on sapphire and AlN/SiC substrates.

Fig. 6.  Comparison of rate of Mg loss during heat-cycle and post-anneal processes to relevant evaporation rates.

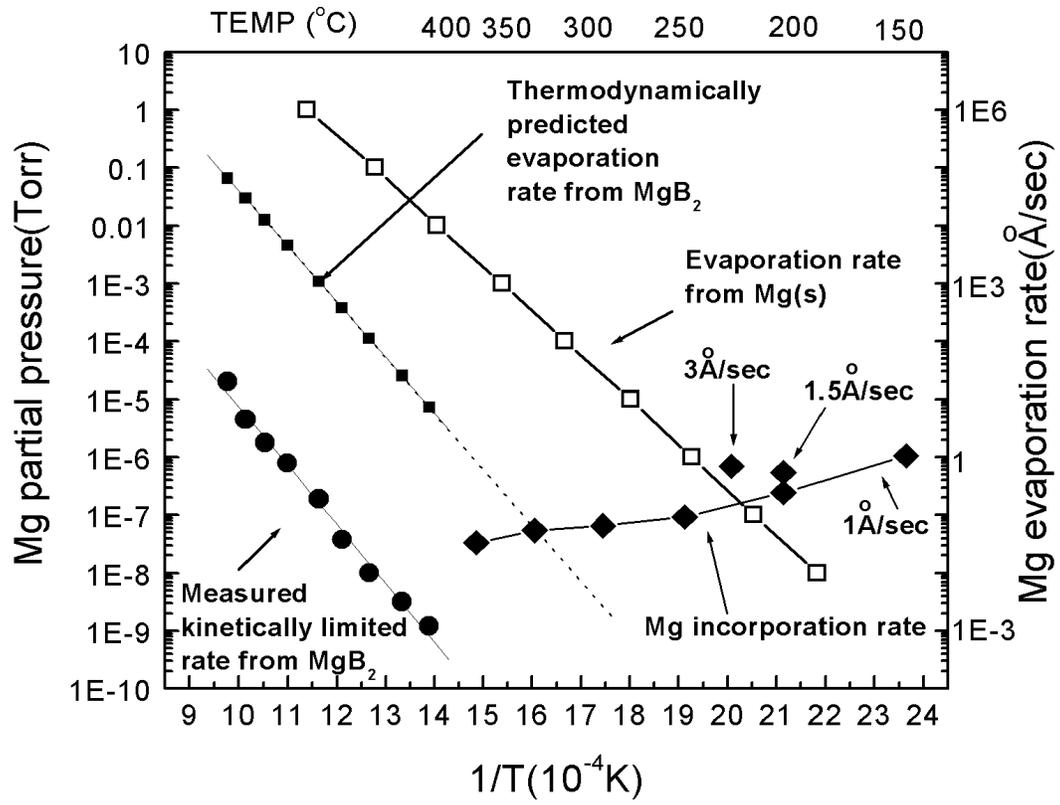

Fig. 1.

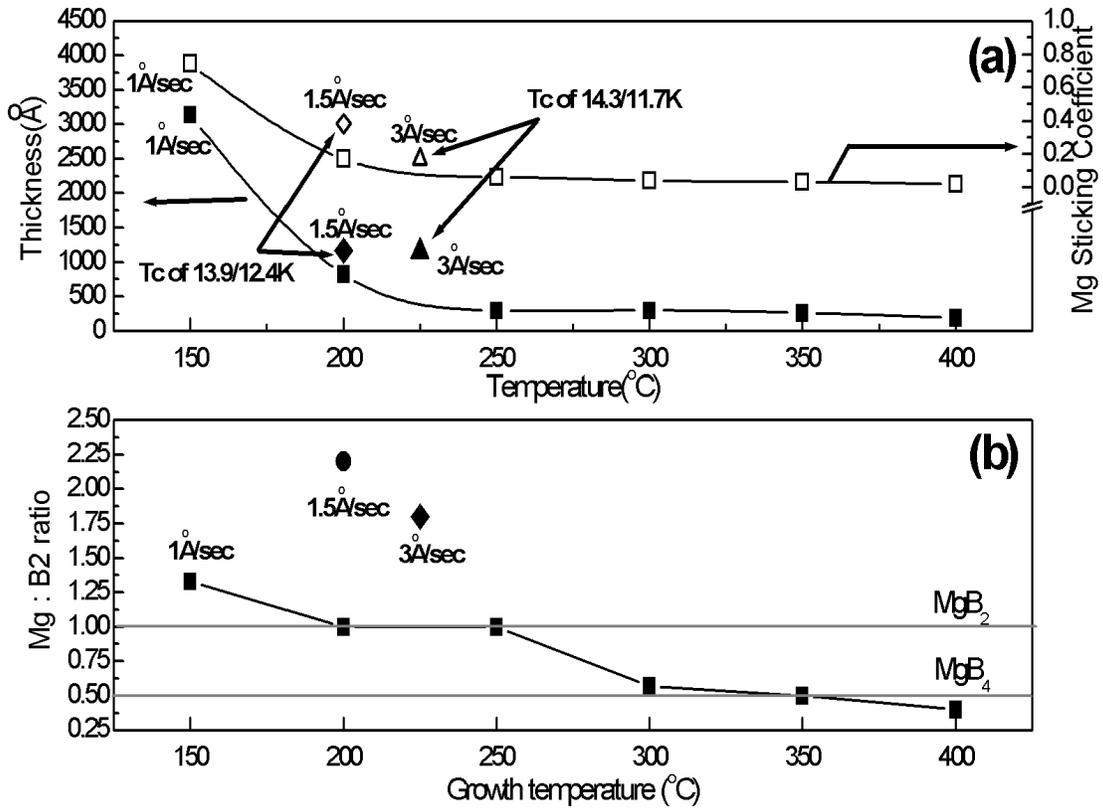

Fig. 2.

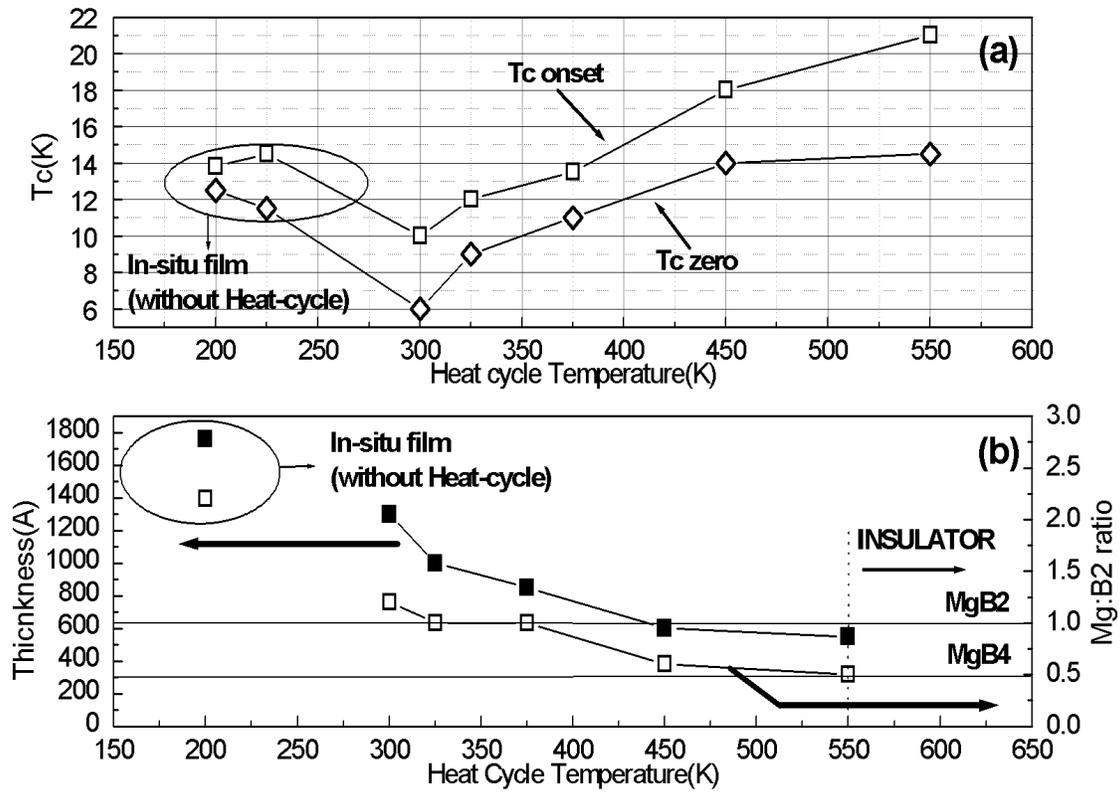

Fig. 3.

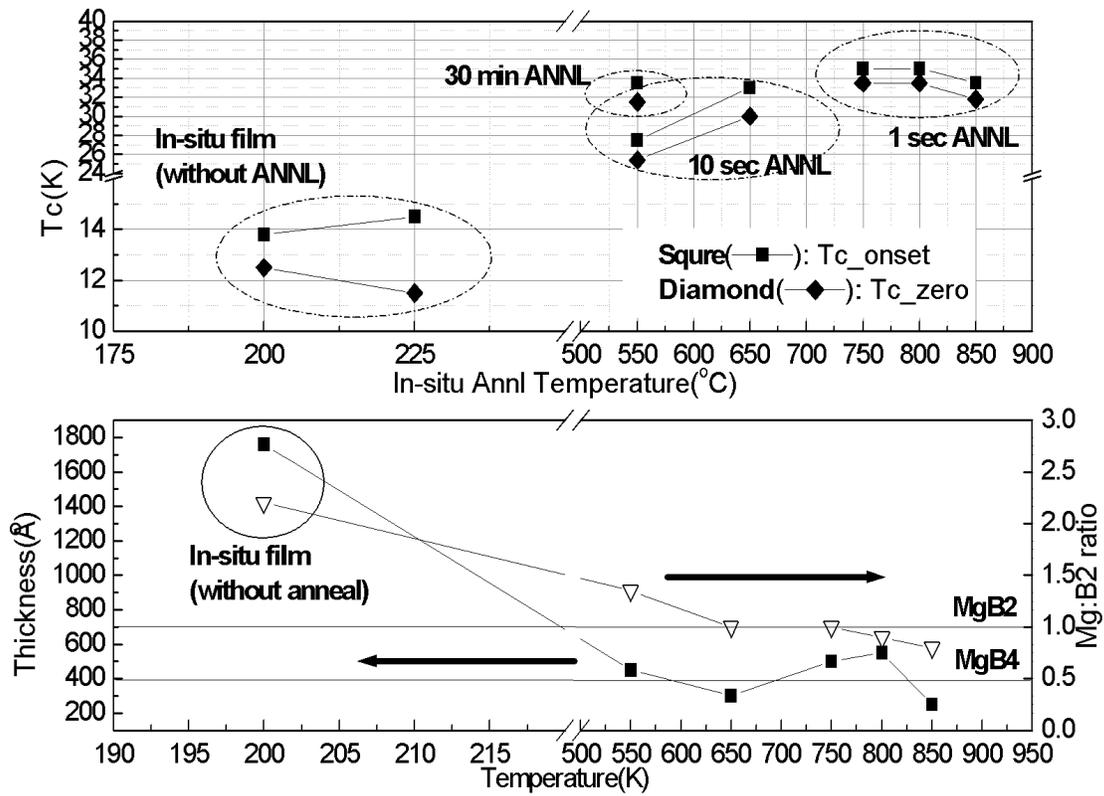

Fig. 4.

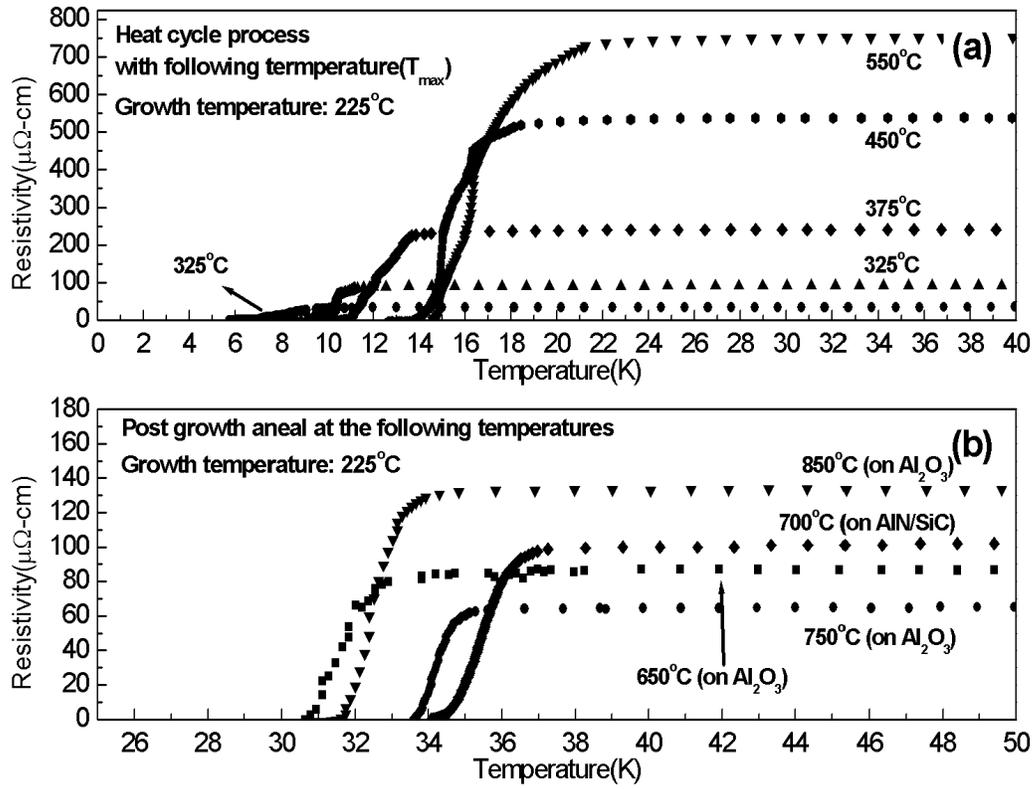

Fig. 5.

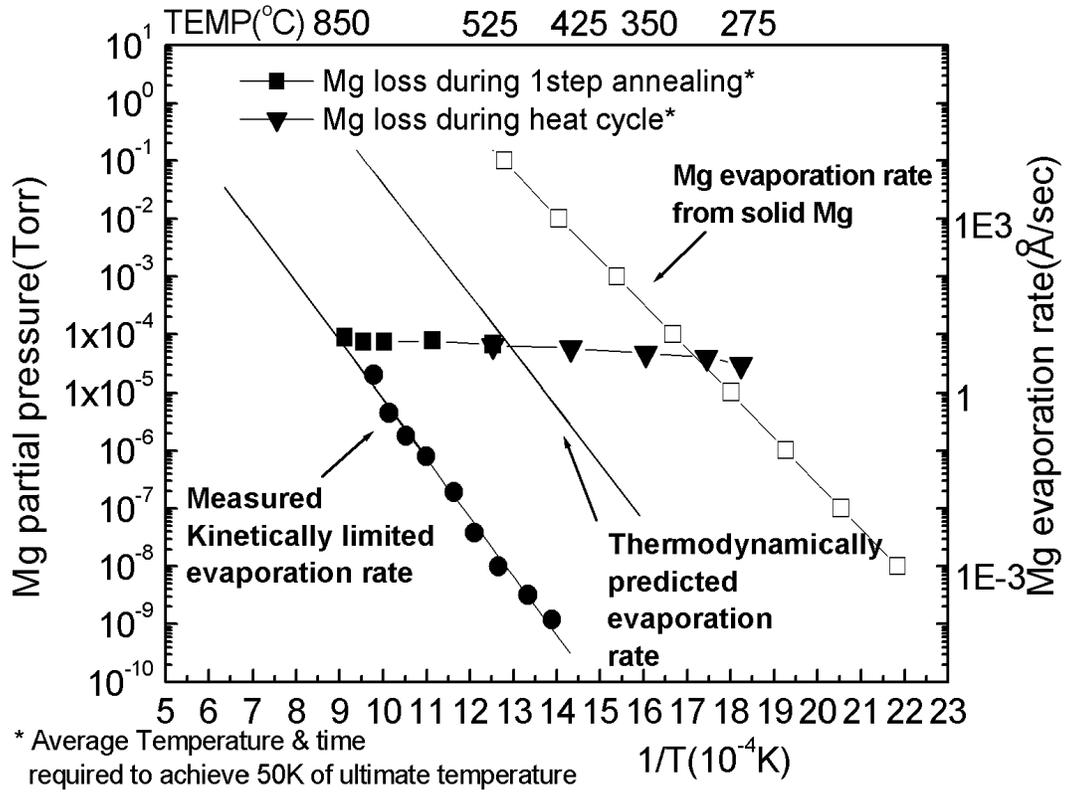

Fig. 6.